 \documentclass[prl,twocolumn]{revtex4}
 \usepackage{amsmath} 
 
\begin{document}
 
 \title{Repulsive Casimir Pistons}
 
\author{S. A. Fulling}
\email{fulling@math.tamu.edu}
 \homepage{http://www.math.tamu.edu/~fulling}
 \affiliation{Departments of Mathematics and Physics, Texas A\&M 
University, 
  College Station, TX, 77843-3368 USA}

  \author{J. H. Wilson}
 \email{thequark@tamu.edu}
\affiliation{Departments of Mathematics and Physics, Texas A\&M 
University, 
  College Station, TX, 77843-3368 USA}

% \date{\today}
 \date{January 7, 2007}

\begin{abstract}
 Casimir pistons are models in which finite Casimir forces can be 
calculated without any suspect renormalizations.
 It has been suggested that such forces are always attractive.
 We present three scenarios in which that is not true.
 Two of these depend on mixing two types of boundary conditions.
 The other, however, is a simple type of quantum graph in which the 
sign of the force depends upon the number of edges.
 \end{abstract} 
 
 \maketitle

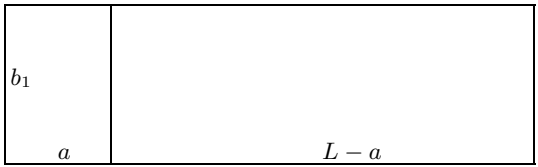
\begin{figure}
\begin{picture}(200,60)
 \put(0,0){\framebox(200,60){}}
 \put(40,0){\line(0,1){60}}
 \put(2,30){$b_1$}
 \put(20,2){$a$}
 \put(120,2){$L-a$}  
\end{picture}
 \caption{A rectangular piston in 
dimension~2 (cf.~\cite{Cav}). 
 In dimension~3   there is another length, $b_2\,$, perpendicular 
to the plane of the figure.} 
 \label{fig:piston}
 \end{figure}

  According to a classic calculation \cite{Luk}, the Casimir force 
  inside a roughly cubical rectangular parallelepiped is repulsive;
 that is, it tends to expand the box.  
The reasoning leading to this conclusion is open to criticism on 
two related grounds: It ignores the possibility of nontrivial 
vacuum energy in the region outside the box, and it involves 
``renormalization'' in the sense of discarding divergent terms 
associated with the boundary although (unlike the case of 
parallel plates, or any calculation of forces between rigid bodies)
 the geometry of the boundary depends upon the dimensions of the 
box.
 Recently 
 (see also \cite{SS})
 a class of scenarios called ``Casimir pistons'' has been 
introduced to which these objections do not apply.
 The  piston is an idealized plate that is free to move along a 
rectangular shaft, whose length, $L-a$,
  to the right of the piston is taken 
arbitrarily large (Fig.~\ref{fig:piston}).
 Both the external region and the divergent
 (or cutoff-dependent) terms in the internal 
vacuum energy are independent of the piston position, $a$,
 so that a well-defined, finite force on the piston is calculated.
 One finds that this force is always attractive,  both for
  a two-dimensional scalar-field model with the Dirichlet boundary 
  condition~\cite{Cav} and a three-dimensional 
 electromagnetic field with the perfect-conductor 
condition~\cite{HJKS}.

 Barton \cite{Bar} showed that the piston force can be repulsive 
for some (not too small) values of~$a$ if the conducting material 
is replaced by a weakly polarizable dielectric. 
This result is somewhat ironic in that one reason for suspicion of 
repulsive Casimir forces is the belief that the force between 
disjoint bodies of realistically modeled material should be always 
attractive.
 It is easily understood, however, as being due to attraction 
between the piston and the distant part of the shaft.
(It would not exist if the shaft extended a long distance to the 
left of the fixed plate (``baffle'') at $a=0$ as well as to the 
right of the piston.)

 In the present note we observe several situations with idealized 
boundary conditions for which the piston force is unambiguously 
repulsive.
 Although these models are less realistic than that studied in 
\cite{HJKS} (or~\cite{Bar}),
 they do show that a repulsive force is not inevitably an artifact 
of a naive renormalization scheme.
 Our effects are unrelated to that in~\cite{Bar} and do not depend 
on the asymmetry just noted in connection with that paper.

 Throughout, we take $\hbar=1=c$.

 \subsection{One-dimensional piston with mixed boundary conditions}
The first example is already rather well known, in its essence.
 Consider a scalar field quantized on the real line divided into 
three parts by two points at each of which either a Dirichlet or a 
Neumann boundary condition is imposed.
 The contributions of the two infinite 
 (or, better, extremely long) intervals 
to the Casimir force will vanish.
 (As emphasized in~\cite{HJKS}, the force contributed by a long 
shaft is entirely associated with periodic orbits perpendicular to 
the shaft.  In dimension~1 such paths don't exist.)
 Let the length of the central interval be~$a$.
 Then the frequencies of the normal modes are 
 \begin{equation}
 \omega_n = \frac{n\pi}{a} 
 \label{DD} \end{equation}
 for positive (or nonnegative) integer~$n$,
 if the boundaries are both Dirichlet (or both Neumann, 
respectively).
 In those cases the well known calculation yields the attractive 
force
 \begin{equation}
 F \equiv -\, \frac{\partial E}{\partial a} = -\,\frac{\pi}{24a^2}\,.
 \label{DDforce}  \end{equation}
 On the other hand, if one boundary is Dirichlet and the other 
Neumann, then the eigenfrequencies are
 \begin{equation}
 \omega_n = \frac{(2n+1)\pi}{2a} 
 \label{DN} \end{equation}
 and the force comes out to be repulsive:
 \begin{equation}
 F  = +\,\frac{\pi}{48a^2}\,.
 \label{DNforce}  \end{equation}
 
 Here is the calculation leading to~(\ref{DNforce}):
 We regularize the sums by an exponential ultraviolet cutoff.
 (The same answer would be obtained by, for example, a calculation 
with zeta functions.)
 It is most convenient to study a sum whose $t$ derivative 
 at $t=0$ is 
proportional to the total regularized energy, namely,
 \begin{equation}
 T(t) \equiv \sum_{n=0}^\infty e^{-\omega_n t}.
 \label{cyltrace}\end{equation}
 One has
 \begin{align*}
 T(t) &\equiv e^{-\pi t/2a} \sum_{n=0}^\infty e^{-\pi nt/a} \\
 &= \frac{e^{-\pi t/2a}}{1 -e^{-\pi t/a} }
 = \frac1{2\sinh(\pi t /2a)} \\
% &\sim \frac{1 - {\pi t/2a} + \frac12({\pi t/2a})^2 + O(t^3)}
%{ {\pi t/a}  - \frac12 ({\pi t/a})^2 + \frac16({\pi t/a})^3 + 
%O(t^4)} \\
% &\sim \frac a{\pi t}
%  \left[ 1  - \frac1{24}\left(\frac{\pi t}{a}\right)^2 
% +\cdots\right] \\
 &\sim
  \frac a{\pi t} - \frac1{24} \,\frac{\pi t}{a} +O(t^2).
 \end{align*}
 Thus the regularized energy is
 \begin{align}
 E(t) &\equiv -\,\frac12\,\frac{\partial T(t)}{\partial t} \nonumber\\
&  =   \frac  a{2\pi t^2} +\frac{\pi}{48a} + O(t).
 \label{DNenergy}\end{align}
 Discarding the leading, cutoff-dependent term (which is compensated 
 in the force
 by similar terms in the exterior regions, already discarded), and 
letting $t\to0$, we arrive at~(\ref{DNforce}).

 More precisely, if the entire space has length~$L$, then the 
regularized energy of the exterior regions is 
 \[ \frac{L-a}{2\pi t^2} + O(L^{-1}). \]
 The second term is negligible as $L\to\infty$, and the first term 
combines with the first term of (\ref{DNenergy}) to make a term 
independent of~$a$, which, therefore, contributes nothing to the 
force.
 Henceforth we shall not repeat this type of argument every time it 
is needed.

 \subsection{Quantum star graphs}

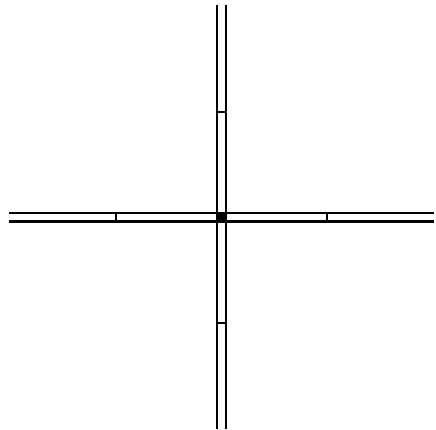
\begin{figure}
 \begin{picture}(160,160)(-80,-80)
 \put(-2,-2){$\bullet$}
 \put(1.6,0){\line(0,1){80}}
  \put(-1.6,0){\line(0,1){80}}
\put(-1.6,40){\line(1,0){3.2}}
 \put(1.6,0){\line(0,-1){80}}
  \put(-1.6,0){\line(0,-1){80}}
\put(-1.6,-40){\line(1,0){3.2}}
 \put(0,1.6){\line(1,0){80}}
  \put(0,-1.6){\line(1,0){80}}
\put(40,-1.6){\line(0,1){3.2}}
 \put(0,1.6){\line(-1,0){80}}
  \put(0,-1.6){\line(-1,0){80}}
\put(-40,1.6){\line(0,-1){3.2}}
\end{picture}
 \caption{A star graph with a piston installed in each edge.
 (The pistons are actually 
points; the edges have no thickness.)}
 \label{fig:star}
 \end{figure}

 In the next model the space consists of $N$ one-dimensional rays of 
length $L$ attached to a central vertex (Fig.~\ref{fig:star}).
 In each ray a Neumann piston is located a distance $a$ from the 
vertex, so that a normal mode of the field in ray~$j$ must take the 
form $u_j(x) = B_j \cos\bigl(\omega(x-a)\bigr)$ when $x$ is 
measured from the center.
 At the central vertex the field has the Kirchhoff 
(generalized Neumann) behavior \cite{Kuc}
 \begin{equation}
 u_j(0) =C  \hbox{ for all $j$}, \quad
 \sum_{j=1}^N u'_j(0)=0 .
 \label{Kirch}\end{equation}
The following analysis is part of a broader study of vacuum energy 
in quantum graphs~\cite{Wil} (see also \cite{Fsb,BM}).

 There are two types of normal modes.
 First, if $\,\cos(\omega a) \ne 0$, we have from (\ref{Kirch})
that  $B_j= C/\cos(\omega a)$ and $\,\tan(\omega a)=0$,
 whence $\omega$ is one of the numbers~(\ref{DD}).
 Second, if $\,\cos(\omega a) = 0$, then $\omega$ 
 is one of the numbers~(\ref{DN}) and 
 \[ \sum_{j=1}^N B_j=0,\]
 which has $N-1$ independent solutions.
 Therefore, the energies and forces are just the appropriate linear 
combinations of those calculated in the previous example:
 the regularized energy for the whole system is
 \begin{equation}
 E(t) = \frac{NL}{2\pi t^2} +\frac{(N-3)\pi}{48a}
 +O(L^{-1}) + O(t), 
 \label{starenergy}\end{equation}
 and the force (either from (\ref{starenergy}) or from 
(\ref{DDforce}) and (\ref{DNforce})) is
 \begin{equation}
 F= \frac{-\pi}{24a^2} + (N-1)\frac{\pi}{48a^2}
 =\frac{(N-3)\pi}{48a^2}\,.
 \label{starforce}\end{equation}

 When $N=1$ or $N=2$, the result reduces properly to that for an 
ordinary Neumann interval of length $a$ or $2a$, respectively.
 When $N>3$, however, the force is repulsive:
if the pistons are free to move, they will tend to move outward.
% (Since we do not have a solution for unequal piston displacements, 
%we cannot exclude the existence of other, asymmetrical modes that 
%are partly attractive.)
 (More generally, a periodic-orbit calculation applicable to 
 unequal piston displacements indicates that the force on each 
individual piston is outward, so there are no other, asymmetrical 
 modes that are partly attractive \cite{Kap}.)
 This effect cannot be attributed to mixed boundary conditions, 
since all the conditions are of the Neumann type.
 (However, replacing all the pistons with Dirichlet pistons while 
maintaining (\ref{Kirch}) would interchange the roles of the two 
eigenvalues and produce attraction for all $N>1$.)

 \subsection{Infinitely permeable piston}
 In principle, a repulsive piston can be constructed in the more 
realistic case of the electromagnetic field in dimension~3,
 in analogy with our original one-dimensional model.
 If the electromagnetic analog of the Dirichlet condition is a 
perfect conductor, then the analog of the Neumann condition is a 
material with infinite magnetic permeability \cite{Boy}.
 (A list of references on this topic appears in \cite{AFG}.)
 The existence of real materials with sufficient permeability to 
exhibit Casimir repulsion in the laboratory is controversial
 \cite{IC,KKMRrep,SZL}.
 Here we merely check that the piston effect discovered by 
Cavalcanti \cite{Cav} and the MIT group \cite{HJKS} does not 
destroy the repulsion shown by less sophisticated calculations.
This is not trivial, since the effect arises from the action of the 
shaft walls on the transverse behavior of the field.

 Following Lukosz \cite{Luk}, but in a notation closer to 
Cavalcanti's (see Fig.~\ref{fig:piston}), we consider a rectangular 
parallelepiped with dimensions $a$, $b_1\,$, and $b_2\,$.
 As previously exemplified, we can calculate a finite vacuum energy 
naively, in full confidence that the discarded divergent terms will 
cancel when a force is calculated for the piston system as a whole.
 We are interested in the case where the piston (the surface that 
is free to move) is infinitely permeable but the shaft and the 
baffle (the rest of the box) are perfect conductors.
 By the Rayleigh--Dowker argument \cite{Dow},
 the energy, $\overline{E}_a\,$, of such a box is
 \begin{equation}
 \overline{E}_a = E_{2a} - E_a\,,
 \label{Rayleigh}\end{equation}
 where $E_a$ is the energy of a totally conducting box also of 
length~$a$.
 By differentiation with respect to~$a$ (not $2a\,$!), this 
relation extends to forces and pressures.
 Thus (\ref{DNforce}) follows from (\ref{DDforce}) by virtue of
 \[ \frac{-\pi} {24a} \left[ \frac12 -1\right] = 
 \frac{-\pi} {24a}  \left[-\,\frac12\right], \]
 and the three-dimensional analogs will involve quantities 
proportional to
\[ \frac1{a^3} \left[\frac18 - 1\right] = 
 \frac1{a^3} \left[-\,\frac78\right]. \]
 
  When $a\ll b_j\,$, Lukosz calculates an attractive pressure
 \[P_a = -\,\frac{\pi^2}{240 a^4}\,,\]
 which implies by (\ref{Rayleigh}) Boyer's formula \cite{Boy}
 \begin {equation}
 \overline{P}_a = +\, \frac78\,  \frac{\pi^2}{240 a^4}
\label{inside}\end{equation}
 for the box with one permeable wall.
 For the opposite limit, $a \gg b_1=b_2=b$ ---
 which we shall need to apply to the external part of the shaft by 
replacing $a$ by $L-a$ --- 
 Lukosz finds a repulsive pressure (involving Catalan's constant)
 \[P = +\, \frac{0.915965}{24b^4}\,.\]
 Just as in \cite{HJKS}, the resulting force is inversely 
proportional to the cross-sectional area and is independent of~$a$, 
 so the corresponding energy term is proportional to~$a$.
 Therefore, application of (\ref{Rayleigh}) gives
 \begin{equation}
 \overline{P}_{L-a} = P_{L-a} =  +\, \frac{0.915965}{24b^4}
 \label{outside}\end{equation}
 (as ought to be the case, since the nature of the plate at the 
distant end of the shaft ought to be irrelevant).
 To find the total force on the piston, we must reverse the sign of 
(\ref{outside}), add it to (\ref{inside}), and multiply by the 
area, $b^2$.
 The point is that the result is positive if $a\ll b$;
 the piston effect is unimportant in that case.
    (Throughout this discussion ``pressure'' simply means ``force 
per area'' without necessarily implying a local pressure 
independent of position on the wall.)

 On the other hand, for a cube Lukosz found that the perfectly 
conducting box was already repulsive.
 The formula (\ref{Rayleigh}) does not yield a simple factor 
 $-\frac12$ in that case, because the doubled box is no longer a 
cube.
 Nevertheless, the graph presented in \cite{HJV}  
 shows that $E_{2a}$ is closer to $\frac12 E_a$ than to $E_a\,$.
 We conclude that 
 the permeable piston is attractive in the cubical 
configuration.

\bigskip
 \begin{acknowledgments}
        We thank Kimball Milton, Lev Kaplan, and the Texas A\&M 
quantum graph research group (Brian Winn, Gregory Berkolaiko, and 
Jonathan Harrison) for helpful comments.  This research is 
supported in part by National Science Foundation Grant No. 
PHY-0554849. 
 \end{acknowledgments}

\goodbreak

\end{document}